\documentclass{elsart}

\usepackage{times}
\usepackage{ae,aecompl}
\usepackage[T1]{fontenc}

\usepackage{graphicx}
\usepackage{amssymb}

\begin{document}

\begin{frontmatter}


\title{Geometrical Phase Transitions}
\author{Wolfhard Janke}
\ead{janke@itp.uni-leipzig.de}
and
\author{Adriaan M. J. Schakel\corauthref{cor1}}
\corauth[cor1]{Corresponding Author}
\ead{schakel@itp.uni-leipzig.de}
\address{Institut f\"ur Theoretische Physik, Universit\"at Leipzig,
Augustusplatz 10/11, 04109 Leipzig, Germany}

\begin{abstract}
  The geometrical approach to phase transitions is illustrated by
  simulating the high-temperature representation of the Ising model on a
  square lattice.

\end{abstract}

\begin{keyword}
geometrical phase transition \sep  plaquette update \sep Ising model
\PACS   02.70.Lq \sep  05.50.+q \sep 75.10.Hk 
\end{keyword}
\end{frontmatter}

\section{Prelude}
\label{sec1} 
The geometrical approach to phase transitions is an exciting research
topic in contemporary physics.  The approach is patterned after
percolation theory which describes clusters of randomly occupied sites
or bonds on a lattice \cite{StauferAharony}.  The fractal structure of
these geometrical objects and whether or not a cluster spans the lattice
are central topics addressed by percolation theory.  By lumping together
with a certain temperature-dependent probability neighboring spins in
the same spin state, spin models such as the $q$-state Potts models can
be mapped onto percolation theory \cite{FK}.  The resulting
Fortuin-Kasteleyn spin clusters percolate at the critical temperature,
while their percolation exponents coincide with the thermal ones.  In
this way, a purely geometrical description of the phase transition in
these models was achieved.

Percolation theory is generic and transparent at the same time, making
it easy to adapt for the description of other geometrical objects such
as lines and domain walls.  Typical line objects featuring in phase
transitions are, for example, (i) vortex lines in superfluids
with a spontaneously broken global U(1) symmetry or in gauge theories,
and (ii) worldlines in Bose-Einstein condensates:

(i) Because of topological constraints, vortices in a superfluid cannot
terminate inside the system and generally form closed loops.  Whereas in
the broken-symmetry phase only a few small vortex loops are present,
loops of all sizes appear at the critical point.  A typical
configuration then has one very large vortex loop and many smaller ones. 
In other words, \textit{the superfluid is pierced through and through
  with vortex line} \cite{Feynman55}.  This vortex proliferation is in
complete analogy to the presence of a spanning cluster at the
percolation threshold in percolation phenomena.  The disordering effect
of the proliferating vortices destroys superfluidity in a superfluid,
and often leads to charge confinement in gauge theories (both Abelian
and non-Abelian). 

(ii) Boson worldlines at finite temperature form closed loops in
imaginary time.  Feynman's theory of Bose-Einstein condensation asserts
that upon lowering the temperature, small loops describing single
particles hook up to form larger exchange rings, so that the particles
become indistinguishable \cite{Feynman53}.  At the critical temperature,
again as in percolation phenomena, worldlines proliferate and loops of
arbitrary size appear, signaling the onset of Bose-Einstein
condensation.  The fractal structure of these worldlines encode the
thermal critical exponents of the phase transition \cite{PRE}.

\begin{figure}
\begin{center}
\includegraphics[width=3.2cm]{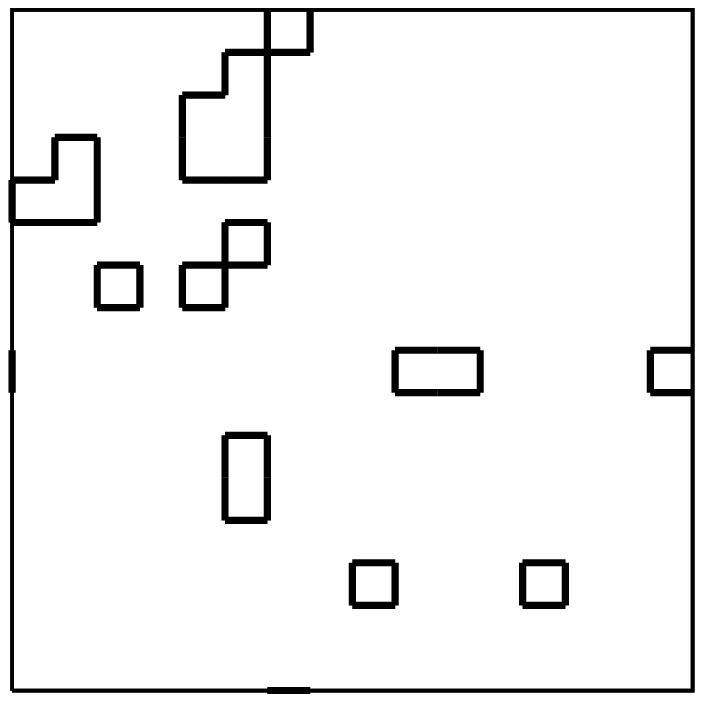} \hspace{0.5cm}
\includegraphics[width=3.2cm]{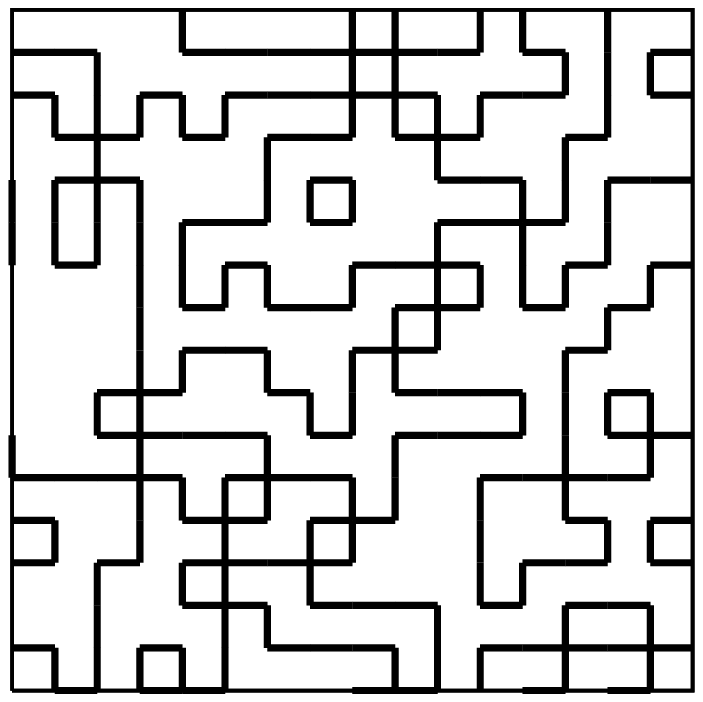} 
\end{center}
\caption{Typical HT graph configurations generated on a $16 \times 16$ square lattice
  with periodic boundary conditions in the high- (left panel) and
  low-temperature (right panel) phase. 
  \label{fig:vis}}
\end{figure}

In this contribution, we report numerical results on the geometrical
approach to the Ising model on a square lattice \cite{geoPotts}.  We
consider the high-temperature (HT) representation of the model which, as
is common for HT or strong-coupling representations, can be represented
by closed graphs on the lattice.  We do not enumerate all possible
graphs to a given order, as is usually done in HT series
expansions \cite{star}, but instead generate HT graphs by means of a
Metropolis plaquette update \cite{Erkinger} and study their fractal
structure (see Fig.~\ref{fig:vis}).  In the high-temperature phase,
large graphs are exponentially suppressed.  Upon lowering the
temperature, graphs of increasing size are generated, cumulating in a
proliferation of graphs at the critical point.  From the percolation
strength $P_\infty$ (defined as the number of bonds per site in the
largest graph) and the average graph size $\chi_\mathrm{G}$, the fractal
dimension of the graphs is extracted through finite-size scaling.  Our
numerical results are in good agreement with the analytic prediction by
Duplantier and Saleur \cite{DS88}, which was derived using the Coulomb
gas map. 
\section{Order and Entropy}
Central in the geometrical description of phase transitions is the
distribution $l_n$ of the geometrical objects under consideration,
\begin{equation} 
l_n \propto n^{-\tau} {\rm e}^{- \theta n},
\end{equation} 
giving the average number density of objects of size $n$ present.  The
distribution consists of two parts.  The second is a Boltzmann factor
which exponentially suppresses large objects.  The suppression
coefficient $\theta$ vanishes with an exponent $1/\sigma$ when the
critical temperature $T_{\rm c}$ is approached, $\theta \propto |T_{\rm
  c} - T|^{1/\sigma}$.  At criticality, only the first factor survives
and the distribution becomes algebraic: $l_n (T_{\rm c}) \propto n^{-
  \tau}$.  This factor, giving the number of ways an object of given
size $n$ can be implemented on the lattice, measures the configurational
entropy.  The exponent $\tau$ is related to the fractal dimension $D$ of
the objects via $\tau = d/D+1$ as in percolation theory, where $d$ is
the space dimension.  The algebraic behavior of the distribution implies
that objects of arbitrary size appear.  Together with the exponent
$\sigma$, the so-called Fisher exponent $\tau$ determines the critical
exponents through scaling relations.  Note that only two independent
exponents are needed to determine the entire set of thermal critical
exponents.  In the geometrical approach, this is reflected by the two
parts comprising the distribution, with both having their own distinct
physical meaning.

The average graph size $\chi_\mathrm{G}$ is given in terms of the graph
distribution $l_n$ as \cite{StauferAharony} $\chi_\mathrm{G} = \sum_n'
n^2 l_n/\sum_n' n l_n$, where the prime on the sum indicates that the
largest graph in each measurement is omitted.  At the critical
temperature, the percolation strength $P_\infty$ and $\chi_\mathrm{G}$
obey the finite-size scaling relations $P_\infty \sim L^{-\beta_{\rm
G}/\nu} , \chi_\mathrm{G} \sim L^{\gamma_{\rm G}/\nu}$, with the graph
exponents \cite{StauferAharony} $ \beta_{\rm G} = (\tau_{\rm G}
-2)/\sigma_{\rm G}, \quad \gamma_{\rm G} = (3-\tau_{\rm G})/\sigma_{\rm
G}$, and $\nu$ the correlation length exponent, which for the 2D Ising
model takes the value $\nu=1$.  Measurement of these two observables
using different lattice sizes gives the two exponents $\beta_{\rm G},
\gamma_{\rm G}$ from which in turn the graph distribution exponents
$\tau_\mathrm{G}$ and $\sigma_\mathrm{G}$ can be extracted. 

\section{Plaquette Update}
\begin{figure}
\begin{center}
\includegraphics[width=5.cm]{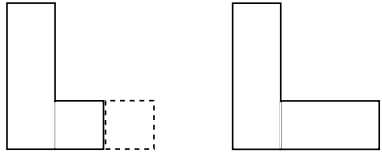}
\end{center}
\caption{Metropolis plaquette update at work. {\it Left panel}: 
Existing HT graph
with the plaquette proposed for updating indicated by the broken square.  
{\it Right panel}: New graph after the proposal is accepted.  
\label{fig:update}}
\end{figure}

The well-known HT representation of the Ising model on a square lattice
reads:
\begin{equation} 
\label{HTIsing}
Z = (\cosh \beta)^{2N} 2^N \sum_{\{\Gamma_{\rm O}\}} v^n,
\end{equation}
where $\{\Gamma_{\rm O}\}$ denotes the set of \textit{closed} graphs
specified by $n$ occupied bonds, $N$ is the total number of sites, and
$v = \tanh \beta$, with $\beta$ the inverse temperature.  The closed
graphs are generated by means of a Metropolis plaquette algorithm, where
a proposed plaquette update resulting in $n'$ occupied bonds is accepted
with probability $p_\mathrm{HT} = \min (1,v^{n'-n})$, with $n$ denoting
the number of occupied bonds before the update \cite{Erkinger}.
Reflecting the Z$_2$ symmetry of the Ising model, all bonds of an
accepted plaquette are changed, i.e., those that were occupied become
unoccupied and \textit{vice versa} (see Fig.~\ref{fig:update}).  By the
famous Kramers-Wannier duality, the HT graphs form Peierls domain walls
between spin clusters of opposite orientation on the dual lattice, and
in the infinite-volume limit the plaquette update is equivalent to a
single spin update on that lattice \cite{geoPotts}.

\section{Results}
\begin{figure}
\begin{center}
\includegraphics[width=7.5cm]{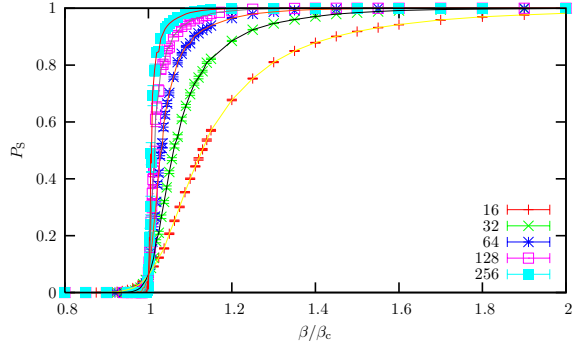} 
\end{center}
\caption{Probability $P_{\rm S}$ for the presence of a spanning graph as
function of the inverse temperature $\beta$ measured for lattice sizes 
$L=16, 32, 64, 128, 256$.  Within the achieved accuracy, the curves cross at the 
thermal critical point $\beta = \beta_{\rm c}$. 
\label{fig:ps}}
\end{figure}

To determine whether the HT graphs proliferate precisely at the critical
temperature, we measure the so-called spanning probability $P_{\rm S}$
as function of $\beta$ for different lattice sizes.  Giving the
probability for the presence of a graph spanning the lattice, $P_{\rm
  S}$ tends to zero for small $\beta$, while it tends to unity for large
$\beta$.  This observable has no scaling dimension and plays the role of
the Binder cumulant in standard thermodynamic studies, so that the
crossing point of the curves obtained for different lattice sizes marks
the proliferation temperature of the infinite system.  Within the
achieved accuracy, we found that the measured curves cross at the
thermal critical point, implying that the HT graphs (domain walls) lose
their line tension $\theta$ and proliferate precisely at the Curie point
(see Fig.~\ref{fig:ps}).  For the graph exponents we found
\cite{geoPotts} $\beta_{\rm G} = 0.626(7), \gamma_{\rm G} = 0.740(4)$,
leading to $\sigma_\mathrm{G} = 0.732(6), \tau_\mathrm{G} = 2.458(5)$ in
perfect agreement with the exact values $\sigma_{\rm G} = 8/11 = 0.7273
\dots, \quad \tau_{\rm G} = 27/11 = 2.4546 \dots$, and the predicted
fractal dimension \cite{DS88} $D_{\rm G} = 11/8$ of the HT graphs.  From
the HT graph exponents all the thermal critical exponents can be
obtained, so that these graphs encode the critical behavior.

\vspace{.8cm}

\noindent 
\textbf{Acknowledgments}

\vspace{.8cm}

This work is partially supported by the DFG grant No. JA 483/17-3 and by
the German-Israel Foundation (GIF) under grant No.\ I-653-181.14/1999. 
A.S. gratefully acknowledges support by the DFG through the
Graduiertenkolleg ``Quantenfeldtheorie'' and the Theoretical Sciences
Center (NTZ) of the Universit\"at Leipzig.  The project is carried out
on a Beowulf GNU/Linux computer cluster. 

\bibliography{ccp04}

\end{document}